\magnification1200


\vskip 2cm
\centerline
{\bf  Generalised geometry, eleven dimensions and $E_{11}$  }
\vskip 1cm
\centerline{ Peter West}
\centerline{Department of Mathematics}
\centerline{King's College, London WC2R 2LS, UK}
\vskip 2cm
\leftline{\sl Abstract}
We construct the non-linear realisation of $E_{11}$ and its first
fundamental representation in eleven  dimensions at low levels. The
fields depend on the usual coordinates of space-time as well as two
form and five form coordinates. We derive the terms in the dynamics that
contain the three form and six form fields and show that when  we
restrict their field dependence to be only on the usual space-time
we recover the correct self-duality relation. Should this result
generalise to the gravity fields then the non-linear realisation is an
extension of the maximal supergravity theory, as previously
conjectured. We also comment on the connections between the different
approaches to generalised geometry.

\vskip2cm
\noindent

\vskip .5cm

\vfill
\eject

\medskip 
{\bf 1. Introduction}
\medskip
It has been realised for more than a quarter of a century that there 
exists no fundamental theory of strings. The maximal supergravity
theories are the complete low energy effective actions for the type II
string theories and these have provided a source of certainty. However,
the study of these theories  has also led to the realisation that a
fundamental theory of strings must include branes, but we have really no
idea how to describe many of the properties of branes. One way forward
may be to try to guess the symmetries of the underlying theory. Some time
ago,  and with  this approach in mind,   it was conjectured that a
non-linear realisation of the Kac-Moody algebra $E_{11}$ was an extension
of the maximal supergravity theories [1]. The different maximal theories
emerge from the different possible decompositions of $E_{11}$ into the
sub-algebras that arise from deleting the  different possible nodes in
the $E_{11}$ Dynkin diagram [2,3,4,5,6,7].  In the early papers on
$E_{11}$ space-time was introduced into  the non-linear realisation by
adjoining to the
$E_{11}$ algebra the space-time translations. It was understood at the
time that this was an adhoc step and in 2003 it was proposed that one
should consider the non-linear realisation of the semi-direct product of
$E_{11}$ and it first fundamental representation $l_1$ denoted
$E_{11}\otimes_s l_1$ [8]. The latter contains the generators $P_a$,
$Z^{ab}$, $Z^{a_1\ldots a_5}$,  $Z^{a_1\ldots a_7, b}$ as well as an 
infinite number of other objects.  In the non-linear realisation this
leads to a generalised space-time with coordinates $x^a$, $x_{ab}$,
$x_{a_1\ldots a_5}$,  $x_{a_1\ldots a_7, b},\ldots $ [8]. There is very
good evidence that the $l_1$ representation contains all the brane
charges [8,9,10,11,3] and so there is a one to one relationship between
the coordinates of the generalised space-time and brane charges. To
appreciate  this proposal one has to understand what is a non-linear
realisation which is in this case not the same as what is often called a
sigma model. Such non-linear realisations were given in the
 papers on $E_{11}$, for example [12,1,5],   and an  early application
was to formulate gravity as a non-linear realisation [14]. The non-linear
realisation of $E_{11}\otimes_s l_1$ not only introduces a generalised
space-time but also a generalised vielbein and so a corresponding
geometry. In particular  this   non-linear realisation   was used to
derive almost all of the features of  the  five dimensional gauged
maximal supergravities [5].  However, there has not been a systematic
attempt to  construct the non-linear realisation of $E_{11}\otimes_s
l_1$. In this paper we will construct the dynamics of the $E_{11}\otimes_s
l_1$ at lowest level keeping the first few coordinates of the
generalised space-time, that is the coordinates $x^a$, $x_{ab}$ and
$x_{a_1\ldots a_5}$ and  the three form and six form fields.

\medskip 
{\bf 2. A review of $E_{11}$ and its $l_1$ representation}
\medskip

In this section we review some of the technical aspects of $E_{11}$ and its first fundamental representation $l_1$ which will be required to construct the non-linear realisation and so  the dynamics. The $E_{11}$ algebra, like any Kac-Moody algebra, is just the multiple
commutator of the Chevalley generators subject to the Serre relations.
 The Dynkin diagram of $E_{11}$ is given by 
$$
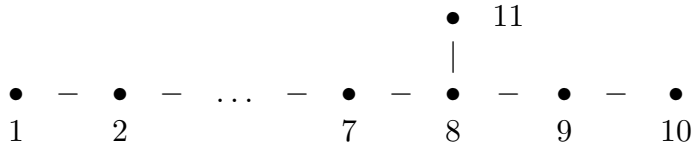
\matrix{
& & & & &&& &\bullet &11&&&
\cr & & & &&& & &| & && &
\cr
\bullet&-&\bullet&-&\ldots &- &\bullet&-&\bullet&-&\bullet&-&\bullet
\cr
1& &2& & & &7& &8& & 9&
&10\cr}
$$
The eleven dimensional theory emerges from  the $E_{11}$ non-linear
realisation  if we delete node eleven and decompose  the $E_{11}$ algebra
in terms of the remaining 
$A_{10}$ subalgebra;  that is decompose the adjoint representation of 
$E_{11}$  in terms representations of $A_{10}$. The generators  can be listed  according to a  level and those of positive level  are given  by [1,13]
$$
K^a{}_b, R^{a_1a_2a_3}, R^{a_1a_2\dots a_6} \ {\rm and }\  R^{a_1a_2\ldots a_8,b}
\eqno(2.1)$$
 at levels zero, one, two and three
respectively.   The generators at level zero  are  those of GL(11)
 and are responsible in the non-linear realisation for eleven dimensional gravity. 
The generator
$R^{a_1a_2\ldots a_8,b}$ obeys the condition $R^{[a_1a_2\ldots a_8,b
]}=0$.  
\par
The corresponding negative level generators  are  given by 
$$ 
R_{a_1a_2a_3}, R_{a_1a_2\dots a_6} \ {\rm and }\ R_{a_1a_2\ldots a_8,b},
\eqno(2.2)$$
at levels -1, -2,-3 with the last generator satisfying an
analogous constraint.  
\par
From the mathematical viewpoint the $E_{11}$ algebra is just the
multiple commutators of the Chevalley generators subject to the Serre
relations.  However, it turns out that the Chevalley generators are
contained in the generators $K^a{}_b$, $R^{a_1a_2a_3}$ and
$R_{a_1a_2a_3}$ and so the $E_{11}$ algebra is found by taking the
multiple commutators of these generators and at low levels it suffices to
just  impose the Jacobi identities on the algebra formed from the
generators listed above. The commutators must  preserve the level and so 
on the right-hand side of the commutators we can only  write
 all possible terms that preserve the level. We can then implement
the Jacobi identities. The level is plus (minus) the number  of times the
positive (negative) root Chevalley generators associated with node eleven
occur in the multiple commutator that creates the generator. However,
this is just the same as plus ( minus) the number of times the generator
$R^{a_1a_2a_3}$ ($R_{a_1a_2a_3}$) occurs in the multiple commutator. 

The generators of GL(11) obey the algebra 
$$
[K^a{}_b,K^c{}_d]=\delta _b^c K^a{}_d - \delta _d^a K^c{}_b,  
\eqno(2.3)$$
By construction the generators in equations (2.1) and (2.2) are
representations of GL(11) and so their  commutators with the   $K^a{}_b$ 
generators are given by 
$$  [K^a{}_b, R^{c_1\ldots c_6}]= 
\delta _b^{c_1}R^{ac_2\ldots c_6}+\dots, \  
 [K^a{}_b, R^{c_1\ldots c_3}]= \delta _b^{c_1}R^{a c_2 c_3}+\dots,
\eqno(2.4)$$
$$ [ K^a{}_b,  R^{c_1\ldots c_8, d} ]= 
(\delta ^{c_1}_b R^{a c_2\ldots c_8, d} +\cdots) + \delta _b^d
R^{c_1\ldots c_8, a} .
\eqno(2.5)$$
where $+\ldots $ means the appropriate anti-symmetrisation. 
The corresponding relations for the negative level generators are 
$$
[K^a{}_b, R_{c_1\ldots c_3}]= -\delta ^a_{c_1}R_{b c_2
c_3}-\dots,\ [K^a{}_b, R_{c_1\ldots c_6}]=  -\delta ^a_{c_1}R_{bc_2\ldots
c_6}-\dots,
\eqno(2.6)$$
$$ [ K^a{}_b,  R_{c_1\ldots c_8, d} ]= 
-(\delta ^a_{c_1} R_{b c_2\ldots c_8, d} +\cdots) - \delta ^a_d
R_{c_1\ldots c_8, b} .
\eqno(2.7)$$
\par
The rest of the $E_{11}$ algebra can be found by remembering that the commutators preserve the level, writing the most general possibility on the right hand side of the commutator, and enforcing the Jacobi identities. 
For the positive level generators we find that [1]
$$[ R^{c_1\ldots c_3}, R^{c_4\ldots c_6}]= 2 R^{c_1\ldots c_6},\ 
\ \ 
[R^{a_1\ldots a_6}, R^{b_1\ldots b_3}]
= 3  R^{a_1\ldots a_6 [b_1 b_2,b_3]}, 
\eqno(2.8)$$
and for the negative root generators 
$$[ R_{c_1\ldots c_3}, R_{c_4\ldots c_6}]= 2 R_{c_1\ldots c_6},\ 
\ \ 
[R_{a_1\ldots a_6}, R_{b_1\ldots b_3}]
= 3  R_{a_1\ldots a_6 [b_1 b_2,b_3]}, 
\eqno(2.9)$$
Finally, the commutation relations between the positive and negative
generators of  up  to level four   are given by [8]

$$[ R^{a_1\ldots a_3}, R_{b_1\ldots b_3}]= 18 \delta^{[a_1a_2}_{[b_1b_2}
K^{a_3]}{}_{b_3]}-2\delta^{a_1a_2 a_3}_{b_1b_2 b_3} D,\  
[ R_{b_1\ldots b_3}, R^{a_1\ldots a_6}]= {5!\over 2}
\delta^{[a_1a_2a_3}_{b_1b_2b_3}R^{a_4a_5a_6]}
$$
$$
[ R^{a_1\ldots a_6}, R_{b_1\ldots b_6}]= -5!.3.3
\delta^{[a_1\ldots a_5}_{[b_1\ldots b_5}
K^{a_6]}{}_{b_6]}+5!\delta^{a_1\ldots  a_6}_{b_1\ldots  b_6} D ,\ \ 
$$
$$
[ R_{a_1\ldots a_3}, R^{b_1\ldots b_8,c}]= 8.7.2
( \delta_{[a_1a_2 a_3}^{[b_1b_2b_3} R^{b_4\ldots b_8] c}-
 \delta_{[a_1a_2 a_3}^{[b_1b_2 |c|} R^{b_3\ldots b_8]} )
$$
$$
[ R_{a_1\ldots a_6}, R^{b_1\ldots b_8,c}]= {7! .2\over 3}
( \delta_{[a_1\ldots  a_6}^{[b_1\dots b_6} R^{b_7 b_8] c}-
 \delta_{[a_1\ldots  a_6}^{c[b_1\ldots b_5 } R^{b_6b_7 b_8]})
\eqno(2.10)$$
where $D=\sum_b K^b{}_b$, $\delta^{a_1a_2}_{b_1b_2}=
{1\over
2}(\delta^{a_1}_{b_1}\delta^{a_2}_{b_2}-
\delta^{a_2}_{b_1}\delta^{a_1}_{b_2})=
\delta^{[a_1}_{b_1}\delta^{a_2]}_{b_2}$ with similar formulae when 
more indices are involved. 
\par
A non-linear realisation is defined by a choice of an algebra  together
with a  subalgebra. For us the subalgebra is chosen to be the one that
 is invariant under the Cartan involution $I_c$. This  is an  
involution, that is
$I_c^2$ is the identity operator,  and an  automorphism  of the algebra,
that is 
$I_c(AB)= I_c(A) I_c(B)$, which   acts  on the generators given above as 
$$
I_c(K^a{}_b) = -\eta ^{ac}\eta_{bd} K^d{}_c,\ I_c(R^{a_1a_2a_3}) = -\eta ^{a_1 b_1}\eta ^{a_2 b_2}\eta ^{a_3 b_3} R_{b_1b_2b_3},
$$
$$
 I_c(R^{a_1\ldots a_6}) = \eta ^{a_1 b_1}\ldots \eta ^{a_6 b_6} 
R_{b_1\ldots b_6},
I_c(R^{a_1\ldots a_8,c}) =
-\eta ^{a_1 b_1}\ldots \eta ^{a_6 b_6} \eta ^{c d}R_{b_1\ldots b_8,d}
\eqno(2.11)$$
Its more fundamental definition in terms of its action on the Chevalley generators can be found in for example [1]. In fact,  we have modified the usual definition to take account of the Minkowski rather than the Euclidean signature. The sub-algebra invariant under the Cartan involution  is generated at low levels by   
$$
J_{ab}=K^c{}_b\eta_{ac}-K^c{}_a\eta_{bc},\ 
S_{a_1a_2a_3}=R^{b_1b_2b_3}\eta_{b_1a_1}\eta_{b_2a_2}\eta_{b_3a_3}
-R_{a_1a_2a_3},
\eqno(2.12)$$
$$
S_{a_1\ldots a_6}=R^{b_1\ldots b_6}\eta_{b_1a_1}\ldots \eta_{b_6a_6}
+R_{a_1\ldots a_6}
\eqno(2.13)$$
$$
S_{a_1\ldots a_8,c}=R^{b_1\ldots b_8,b}\eta_{b_1a_1}\ldots \eta_{b_8a_8}
\eta_{b c}-R_{a_1\ldots a_8,c} .
\eqno(2.14)$$
where $\eta_{ab}$ is the metric of Minkowski space-time. 
The generators $J_{ab}$ are those of the Lorentz algebra SO(1,10) and
their commutators with   the other generators just express the fact that
they belong to a   representation of the Lorentz algebra. 
 The $S_{a_1a_2a_3}$  and $S_{a_1\ldots a_6}$
generators obey the commutators [8]
$$
[ S^{a_1a_2a_3}, S_{b_1b_2b_3}]=-18\delta^{[a_1a_2}_{[b_1b_2}
J^{a_3]}{}_{b_3]}+2S^{a_1a_2a_3}{}_{b_1b_2b_3}
\eqno(2.15)$$
$$
[ S_{a_1a_2a_3}, S^{b_1\ldots b_6}]=-{5!\over
2}\delta^{[b_1b_2b_3}_{a_1a_2a_3} S^{b_4b_5b_6]}-3S^{b_1\ldots
b_6}{}_{[a_1a_2,a_3]}
\eqno(2.16)$$
\par
The non-linear realisation of interest to us also includes generators in the fundamental representation of
$E_{11}$, denoted $l_1$;  by definition this representation  has highest weight $\Lambda_1$ which  obeys 
$(\Lambda_1, \alpha_a )=\delta _{a,1}$. Decomposed into representations of GL(11) the $l_1$ representation contains, at  low levels,   the
generators  [8,9]
$$
P_a, Z^{ab}, Z^{a_1\ldots a_5}, Z^{a_1\ldots a_7,b}, Z^{a_1\ldots a_8},
Z^{b_1 b_2 b_3,a_1 ...a_8}, \ldots  
\eqno(2.17)$$
at levels 0,1,2,3,3 and 4 respectively. Here $P_a$ is the generator of
space-time translations and the next two generators can be identified
with the central charges in the supersymmetry algebra. 
The commutators of the low level generators of the
$l_1$ representation with $R^{a_1a_2a_3}$ are determined up to constants
by demanding that the levels match and so we can take [8]
$$
[R^{a_1a_2a_3}, P_b]= 3 \delta^{[a_1}_b Z^{a_2a_3]}, \ 
[R^{a_1a_2a_3}, Z^{b_1b_2} ]= Z^{a_1a_2a_3 b_1b_2},\ 
$$
$$[R^{a_1a_2a_3}, Z^{b_1\ldots b_5} ]=Z^{b_1\ldots b_5[a_1a_2,a_3]}+
Z^{b_1\ldots b_5 a_1a_2 a_3}
\eqno(2.18)$$ These equations define the normalisation
of these generators of the $l_1$ representation. 
\par
We will be interested in constructing the non-linear realisation of the
semi-direct product of the $E_{11}$ algebra with its $l_1$
representation, denoted $E_{11}\otimes_s l_1$.  In this algebra the
commutators of the generators of
$E_{11}$ with themselves obey the same algebra as above that is equations
(2.4) to (2.10). We will take the generators of the $l_1$ representation
to commute with themselves. The commutators between the generators of
$E_{11}$ and those of the $l_1$ representation express the fact that they
are a representation of $E_{11}$ and this is enforced by demanding the
Jacobi identity involving two $E_{11}$ generators and one $l_1$
generator. The construction is essentially the same as that for the
Poincare group where the Lorentz group L plays the role of $E_{11}$ and
the space-time translations $T$ the role of the $l_1$ representation;
that is we can write the Poincare group as $L\otimes_s T$. 
\par
As we have decomposed the $l_1$ representation into representations of
GL(11),  the commutators of these generators with those of the $l_1$
representations are given [8]
$$
 [K^a{}_b, P_c]= -\delta _c^a P_b +{1\over
2}\delta _b^a P_c,\  [K^a{}_b, Z^{c_1c_2} ]= 2\delta_b^{[c_1} Z^{|a|c_2]}
+{1\over 2}\delta _b^a Z^{c_1c_2},\ 
$$
$$
[K^a{}_b, Z^{c_1\ldots c_5} ]= 5\delta_b^{[c_1} Z^{|a|c_2\ldots c_5]}
+{1\over 2}\delta _b^a Z^{c_1\ldots c_5}
\eqno(2.19)$$
The term with the factor of ${1\over 2}$ plays an important role in many 
applications of $E_{11}$ and it follows from the fact that the $l_1$ is a
highest weight representation of $E_{11}$ [8]. Strictly speaking it is
actually a  lowest weight representation as usually defined.  We
also find using the Jacobi identities and equations (2.8) and (2.18) 
that 
$$
[R^{a_1\dots a_6}, P_b]= -3 \delta^{[a_1}_b Z^{\ldots a_6]}, \ 
[R^{a_1\dots a_6}, Z^{b_1b_2} ]= -Z^{b_1b_2[a_1\ldots
a_5,a_6]}-Z^{b_1b_2 a_1\ldots a_6},\ 
\eqno(2.20)$$
The  commutators with the negative root generators are given
by 
$$
[R_{a_1a_2a_3}, P_b ]= 0,\ 
[R_{a_1a_2a_3}, Z^{b_1b_2} ]= 6\delta^{b_1b_2 }_{[a_1a_2} P_{a_3 ]},\ 
[R_{a_1a_2a_3}, Z^{b_1\ldots b_5} ]= {5!\over 2} \delta^{[ b_1b_2b_3
}_{a_1a_2a_3} Z^{b_4b_5]}
\eqno(2.21)$$
The first equation, just follows from the fact that the $l_1$ representation is a highest weight representation.  and   the subsequent equations follow by  using the
equation  (2.10) and the  Jacobi identities.  
\par
To conclude this section we  now also give the commutation relations between the generators of the Cartan involution
invariant subalgebra,  given in equations (2.12-2.14),  and the 
generators of the $l_1$ representation [8] 
$$
[S^{a_1a_2a_3}, P_b ]= 3\delta _b^{[a_1}Z^{a_1a_3]},\ 
[S_{a_1a_2a_3}, Z^{b_1b_2} ]=Z_ {a_1a_2a_3}{}^{b_1b_2}
-6\delta^{b_1b_2}_{[a_1a_2 }P_{a_3]},\ \  
$$
$$
[S_{a_1a_2a_3}, Z^{b_1\ldots b_5} ]= Z^{b_1\ldots b_5}{}_{[a_1a_2,a_3]}
+Z^{b_1\ldots b_5}{}_{ a_1a_2 a_3}
-{5!\over 2} \delta^{[b_1\ldots b_3}_{a_1a_2a_3}Z^{b_4b_5]}
\eqno(2.24)$$

\medskip
{\bf 3.  The non-linear realisation of $E_{11}\otimes_s l_1$}
\medskip
It was conjectured [8] that the non-linear realisation of
$E_{11}\otimes_sl_1$ was an extension of the equations of motion of 
eleven dimensional supergravity. Put another way, it states that eleven
dimensional supergravity was contained in the
non-linear realisation of $E_{11}\otimes_sl_1$ at low levels. 
At higher levels one finds not only an infinite number of new fields
coming from $E_{11}$, but also all the  fields depend on a generalised
space-time encoded in the $l_1$ representation.  
\par The   non-linear
realisation of
$E_{11}\otimes_sl_1$  is constructed from  a group element
$g\in E_{11}\otimes_sl_1$ which can be written as 
$$
g=g_lg_E 
\eqno(3.1)$$
where 
$$
g_E=  e^{ A^{a_1\ldots a_3} R_{a_1\ldots
a_3}} e^{ A^{a_1\ldots
a_6} R_{a_1\ldots a_6}}e^{ h^{a_1\ldots a_{8},b}R_{a_1\ldots a_{8},b}}\ldots 
$$
$$ e^{h_a{}^b K^a{}_b}\ldots e^{ h_{a_1\ldots a_{8},b}
R^{a_1\ldots a_{8},b}} e^{ A_{a_1\ldots
a_6} R^{a_1\ldots a_6}}e^{ A_{a_1\ldots a_3} R^{a_1\ldots
a_3}}
\eqno(3.2)$$ 
and 
$$
g_l= e^{x^aP_a} e^{x_{ab}Z^{ab}} e^{x_{a_1\ldots a_5}Z^{a_1\ldots a_5}}\ldots = 
e^{z^A L_A} 
\eqno(3.3)$$
where we have denoted the generalised coordinates by $z^A$ and the generators of the $l_1$ representation by $L_A$. Thus the non-linear realisation introduces a generalised space-time with the coordinates [8]
$$
x^a, x_{ab}, x_{a_1\ldots a_5}, \ldots 
\eqno(3.4)$$
The fields that occur in the group element $g_E$ are taken to depend on
the generalised space-time that is the coordinates of equation (3.4).
Since the $l_1$ representations contains all the brane charges
[8,9,3,11] and this was responsible for the generalised space-time there
is a one to one relation between the brane charges and the coordinates of
the generalised space-time.  Furthermore,  for every field in $E_{11}$ 
there corresponds an element in the $l_1$ representation [9]. As such for
every field there is an associated coordinate in the generalised
space-time. For example, the metric corresponds to the point particle
with charge
$P_a$ and coordinate $x^a$, the three form corresponds to the two brane
with charge $Z^{a_1a_2}$ and coordinate $x_{a_1a_2}$, the six form
corresponds to the five brane with charge $Z^{a_1\ldots a_5}$ and
coordinate $x_{a_1\dots a_5}$ and so on. As the discussion below makes
clear, the non-linear realisation $E_{11}\otimes_sl_1$ automatically
encodes a generalised geometry equipped with a generalised vielbein. 
\par
The non-linear realisation is by definition  just a set of dynamics that
is invariant under the transformations 
$$
g\to g_0 g, \ \ \ g_0\in E_{11}\otimes _s l_1,\ \ {\rm as \  well \  as} \ \ \ g\to gh, \ \ \ h\in
I_c(E_{11})
\eqno(3.5)$$
The group element $g_0$ is a rigid transformation, that is a constant,  while $h$ is a local
transformation, that is it depends on the generalised space-time. 
As the generators in $g_l$ form a representation of $E_{11}$ the above transformations for $g_0\in E_{11}$ can be written as 
$$
g_l\to g_0 g_lg_0^{-1}, g_E\to g_0 g_E\quad {\rm and } \quad g_E\to g_E h
\eqno(3.6)$$
As a consequence the coordinates are inert under the local transformations but transform under the rigid  transformations as 
$$
z^A l_A\to g_0 z^Al_A g_0^{-1} = z^\Pi D(g_0^{-1})_\Pi {}^AL_A
\eqno(3.7)$$
\par
Using the local transformation we may bring $g_E$ into the form 
$$
g_E= e^{h_a{}^b K^a{}_b} \ldots e^{ h_{a_1\ldots a_{8},b}
R^{a_1\ldots a_{8},b}} e^{ A_{a_1\ldots
a_6} R^{a_1\ldots a_6}}e^{A_{a_1\ldots a_3} R^{a_1\ldots
a_3}}
\eqno(3.8)$$ 
Thus the theory contains 
the graviton field $h_a{}^b$ associated with the generators $K^a{}_b$
of   $GL(11)$, as well as the gauge fields 
$A_{a_1a_2a_3}$ 
 and its dual
$A_{a_1\ldots a_6}$ associated with  the level one and two generators 
$R^{a_1a_2a_3}$  and
$R^{a_1\ldots a_6}$ respectively. Furthermore, in addition we have a field 
$ h_{a_1\ldots a_8,b}$ corresponding to the generator $R^{a_1\ldots
a_8,b}$ which is the dual field of gravity [1].  The parameterisation of
the group element differs from that used in some earlier works on
$E_{11}$, but this does not affect any physical results. 
\par
The dynamics is usually constructed from the Cartan forms 
${\cal V}=g^{-1}d g$ as these  are inert under the $E_{11}$ rigid
transformations of equation (3.5) and only transform under the local
transformations as 
$$
{\cal V}\to h^{-1} {\cal V} h+ h^{-1} d h
\eqno(3.9)$$
Hence if we use the Cartan forms, the problem of finding a set of field equations which are invariant under equation (3.5) 
 reduces to
finding a set that is invariant under the local subalgebra  $I_c(E_{11})$, that is the transformations of equation (3.9). 
\par
The  Cartan forms  can be written as  
$${\cal V}= {\cal V}_E+{\cal V}_l
\eqno(3.10)$$
where 
$$
{\cal V}_E=g_E^{-1}dg_E \quad {\rm and }\quad {\cal V}_l= g_E^{-1}(g_l^{-1}dg_l) g_E
\eqno(3.11)$$
The first part ${\cal V}_E$ is just the Cartan form for $E_{11}$ while ${\cal V}_l$ is a sum of generators in the $l_1$ representation. 
While  both ${\cal V}_E$ and ${\cal V}_l$ are invariant under rigid transformations and under local transformations they change as 
$$ 
{\cal V}_E\to h^{-1}{\cal V}_E h + h^{-1} d h\quad {\rm and }\quad 
{\cal V}_l\to h^{-1}{\cal V}_l h 
\eqno(3.12)$$
\par 
Let us  evaluate the $E_{11}$ part of the Cartan form 
$$
{\cal V}_E = dz^\Pi G_{\Pi, \star} R^\star= G _{a}{}^b K^a{}_b +  G_{c_1\ldots c_3} R^{c_1\ldots c_3} +G_{c_1\ldots
c_6} R^{c_1\ldots c_6}+
G_{c_1\ldots c_8,b} R^{c_1\ldots c_8,b}+\ldots 
\eqno(3.13)$$
where  $\star $ denotes the indices on the generators of $E_{11}$. Explicitly one finds that  
$$
G _{a}{}^b=(e^{-1}d e)_a{}^b,\ \ G_{c_1\ldots c_3}=\tilde D
A_{c_1\ldots c_3}, 
$$
$$ 
G_{c_1\ldots c_6}=  \tilde D A_{c_1\ldots c_6} 
- A_{[ c_1\ldots c_3}\tilde D A_{c_4\ldots c_6]}
$$
$$
G_{c_1\ldots c_8,b} = \tilde D  h_{c_1\ldots c_8,b}
-A_{[c_1\ldots c_3}\tilde D A_{c_4c_5c_6} A_{c_7c_8 ]b}
+3 A_{[c_1\ldots c_6}\tilde D A_{c_7c_8 ]b}
$$
$$
+(A_{[c_1\ldots c_3}\tilde D A_{c_4c_5c_6} A_{c_7c_8  b]}
-3 A_{[c_1\ldots c_6}\tilde D A_{c_7c_8 b]})
\eqno(3.14)$$
 where $e_\mu{}^a \equiv (e^h)_\mu{}^a$ and  
$$
\tilde D A_{c_1\ldots c_3}\equiv  d A_{c_1 c_2 c_3} 
+ ((e^{-1}d e)_{c_1}{}^{b}A_{b c_2 c_3}+\ldots
\eqno(3.15)$$
where $+\ldots $
denotes the action of 
$(e^{-1}d e)$ on the other 
indices with analogous expressions for other quantities. In the last
expression of equation (3.14) we have subtracted the totally
anti-symmetric  part 
corresponding to the fact that the generator obeys the condition 
$R^{[c_1\ldots c_8,b]}=0$.  Evaluating this expression we find that 
$$
G_{c_1\ldots c_8,b} = \tilde D  h_{c_1\ldots c_8,b}
-A_{[c_1\ldots c_3}\tilde D A_{c_4c_5c_6} A_{c_7c_8 ]b}
+2 A_{[c_1\ldots c_6}\tilde D A_{c_7c_8 ]b}
$$
$$
+2\tilde D A_{[c_1\ldots c_5|b |} A_{c_6c_7 c_8]})
\eqno(3.16)$$
We note that 
$$
A_{[c_1\ldots c_3}\tilde D A_{c_4c_5c_6} A_{c_7c_8  b]}=0
\eqno(3.17)$$
\par
Let us now evaluate the part of the Cartan form in equation
(3.10) containing the generators of the $l_1$ representation; we may
write it as  
$$
{\cal V}_l= g^{-1} d g= dz^\Pi E_\Pi{}^A l_A= 
g_E^{-1}(dx^aP_a+ dx_{ab}Z^{ab}+dx_{a_1\ldots a_5} Z^{a_1\ldots a_5}+\ldots ) g_E
\eqno(3.18)$$
Using equation (2.18-21) we find that 
  ${ E}_\Pi{}^A$, viewed as   a matrix,  is given at low orders by 
$$
{ E}= (det e)^{-{1\over 2}}
\left(\matrix {e_\mu{}^a&-3 e_\mu{}^c A_{cb_1b_2}& 3 e_\mu{}^c A_{cb_1\ldots b_5}+{3\over 2} e_\mu{}^c A_{[b_1b_2b_3}A_{|c|b_4b_5]}\cr
0&(e^{-1})_{[b_1}{}^{\mu_1} (e^{-1})_{b_2]}{}^{\mu_2}&- A_{[b_1b_2b_3 }(e^{-1})_{b_4}{}^{\mu_1} (e^{-1})_{b_5 ]}{}^{\mu_2}  \cr
0&0& e^{-1})_{[b_1}{}^{\mu_1} \ldots (e^{-1})_{b_5]}{}^{\mu_5}\cr}\right)
\eqno(3.19)$$
This illustrates the fact that  the  generalised space-time leads to a 
generalised tangent space, which in this case has the usual tangent
space, two forms, five forms and higher objects. In general the tangent
space can be  read off from the $l_1$ representation in an obvious way.
The $l_1$ representation appropriate to ten and $d$ dimensions is found
by decomposing 
$E_{11}$ into $GL(d)\otimes E_{11-d}$ and the results  can be found in
[4,9,11,21]. The tangent space group is $I_c(E_{11})$; at lowest
level this is  $O(d)\otimes
O(d)$  for the IIA theory in ten dimensions while in
$d$ dimensions it is $SO(d)\otimes I_c(E_{11-d})$
\par
Our task is to find a set of dynamics which is  invariant under the rigid
and local transformations of equation (3.5) and with this in mind we now
consider in more detail the transformations of the two parts of the
Cartan form beginning with that of $E_{11}$ part.  As noted above the
Cartan forms only transform  under the local $I_c(E_{11})$
transformations. It is useful to introduce the operation 
$g^*=(I_c(g))^{-1}$ on the group. While $I_c$ is an automorphism, i.e. on two group elements 
$I_c(g_1g_2)=I_c(g_1)I_c(g_2)$, the action of $*$ reverses the order, that is 
$(g_1g_2)^* =(g_2)^* (g_1)^* $. The action of $*$ on  the algebra is given by $A^*=-I_c(A)$ and $(AB)^*=B^*A^*$. A group element belonging to   $I_c(E_{11})$ obeys 
$h^*=h^{-1}$ and  the two transformations of equation (3.5) imply that 
$g^*\to h^{-1}g^* (g_0)^*$.  
We write the Cartan forms ${\cal V}_E$ as 
$$
{\cal V}_E=P+Q, \ \ {\rm where }\ \ P={1\over 2}({\cal V}_E+{\cal
V}_E^*),\ Q={1\over 2}({\cal V}_E-{\cal
V}_E^*)
\eqno(3.20)$$
and  then  the transformations of equation (3.12) becomes  
$$
P\to h^{-1}Ph,\ \ Q\to h^{-1}Qh+  h^{-1} dh
\eqno(3.21)$$
Examining equation (3.13)  we find that 
$$
P= {1\over 2}G _{a}{}^b (K^a{}_b +K_b{}^a)+  {1\over 2}G_{c_1\ldots c_3}
(R^{c_1\ldots c_3}+R_{c_1\ldots c_3}) +{1\over 2}G_{c_1\ldots c_6}
(R^{c_1\ldots c_6}-R_{c_1\ldots c_6})
$$
$$
+ {1\over 2}G_{c_1\ldots c_8,b}
(R^{c_1\ldots c_8,b}+R^{c_1\ldots c_8,b})+\ldots 
\eqno(3.22)$$
and 
$$
Q= {1\over 2}G _{a}{}^b (K^a{}_b -K_b{}^a)+  {1\over 2}G_{c_1\ldots c_3}
(R^{c_1\ldots c_3}-R_{c_1\ldots c_3}) +{1\over 2}G_{c_1\ldots c_6}
(R^{c_1\ldots c_6}+R_{c_1\ldots c_6})
$$
$$
+ {1\over 2}G_{c_1\ldots c_8,b}
(R^{c_1\ldots c_8,b}-R^{c_1\ldots c_8,b})+\ldots 
\eqno(3.23)$$
We note that the connection $Q$ contains the same objects as the
covariant quantity $P$. 
\par
Taking $h=1-\Lambda_{a_1a_2a_3}S^{a_1a_2a_3}$, the local 
transformations of $P$ of equation (3.21) implies, 
using equations (2.12-14) and equations (2.18-21)  that 
$$
\delta G^{ab}=18 \Lambda^{c_1c_2 b }G_{c_1c_2 }{}^{a}
-2 \delta ^{ab}  \Lambda^{c_1c_2 c_3}G_{c_1c_2 c_3},\ 
$$
$$
\delta G_{a_1a_2a_3}=-{5!\over 2} G_{b_1b_2b_3 a_1a_2a_3}
\Lambda^{b_1b_2 b_3}-6G^{c}{}_{[a_1 } \Lambda_{|c|a_2a_3]},\ 
$$
$$
\delta G_{a_1\ldots a_6}=2 \Lambda_{[ a_1a_2a_3}G_{a_4a_5a_6 ]}
-8.7.2 G_{b_1b_2b_3 [ a_1\ldots a_5,a_6]}\Lambda^{b_1b_2b_3}
+8.7.2 G_{b_1b_2[ a_1\ldots a_5a_6, b_3 ]}\Lambda^{b_1b_2b_3}
$$
$$
\delta G_{a_1\ldots a_8,b}=-3 G_{[ a_1\ldots a_6}\Lambda_{a_7a_8] b}
+3 G_{[ a_1\ldots a_6}\Lambda_{a_7a_8 b]}
\eqno(3.24)$$
\par
Let us now turn our attention to the transformation of the part of the
Cartan form in the direction of the $l_1$ representation, that is ${\cal
V}_l$. At lowest level equation (3.12) implies that 
$$
E_\Pi{}^{A\prime}= E_\Pi{}^B D(h)_B{}^A,\quad {\rm and \ for\  the \
inverse }\quad  (E^{-1})_A{}^{\Pi\prime}= D(h^{-1})_A{}^B (E^{-1})_B{}^\Pi
\eqno(3.25)$$
if we define $h^{-1}L_A h= D(h)_A{}^B L_B$.   
 At lowest levels this implies the local transforms 
$$
\delta E_\Pi{}^a= -6E_\Pi{}_{b_1b_2} \Lambda^{b_1b_2 a}, \quad \delta
E_\Pi{}_{a_1a_2}= 3\Lambda_{a_1a_2 b}  E_\Pi{}^b,\ldots 
$$
$$
\delta (E^{-1})_a{}^\Pi= -3(E^{-1})^{b_1b_2}{}^\Pi
\Lambda_{b_1b_2 a},
\quad \delta (E^{-1})^{a_1a_2}{}^\Pi= 6\Lambda^{a_1a_2
b}  (E^{-1})_b{}^\Pi,\ldots 
\eqno(3.26)$$
\par
Even though the Cartan forms are invariant under the rigid transformations,  
 $E_\Pi{}^A$ and $G_{\Pi \star}$ are not as the transformation of $z^\Pi$
of equation (3.7) implies a corresponding inverse transformation acting
on the $\Pi$ index of these two objects. Thus  $E_\Pi{}^A$ transforms
under a local transformation on its $A$ index and by the inverse of the
coordinate transformation on its $\Pi$ index. As such we can think of it
as a generalised vielbein.   We can rewrite the Cartan form of
$E_{11}\otimes_s l_1$ as 
$$
{\cal V}= g^{-1}d g= dz^\Pi E_\Pi{}^A( L_A+Ê G_{A,*} R^*)
\eqno(3.27)$$
where  $G_{A,*}=(E^{-1})_A{}^\PiÊ G_{\Pi,*}$. At low levels $(E^{-1})_A{}^\Pi$ is the inverse of the matrix of equation (3.19). ÊClearly $G_{A,*}$ is inert under rigid transformations, but it  transforms under local transformations as in equation (3.24) on its $\star$ index  and as the inverse generalised vielbein on its $A$ index,  that is as in equation (3.26). 
\par
Thus if  we choose to construct the dynamics out of $G_{A,*}$ we need only
worry about  the local transformations as the rigid transformations are
automatically taken care of. Hence we seek a set of equations that are 
first order in $G_{A,*}$  and invariant under $I_c(E_{11})$
transformations; thus we are left with a problem in group theory.
\par
We will now focus our attention on finding the terms in such an invariant dynamics that involve the three and six form gauge fields and the coordinates $x^a, x_{ab}$ and $x_{a_1\ldots a_5}$. We can solve this problem using a trick which may not generalise to the full system. In [8] it was shown at  the lowest levels  that $I_c(E_{11})$ is the group SL(32) and the generators $P_a, Z^{ab}, Z^{a_1\ldots a_5}$ can be collected together in the matrix 
$$
Z_\alpha{}^\beta= (\gamma^a P_a+ {\gamma_{ab}\over 2} Z^{ab}+ {\gamma_{a_1\ldots a_5}\over 5!}Z^{a_1\ldots a_5})_\alpha{}^\beta
\eqno(3.28)$$
where $\alpha, \beta=1,\ldots , 32$ and the $\gamma^a$ matrices are elements of the  eleven dimensional Clifford algebra. In fact these first few components of the $l_1$ representation are the charges that occur in the eleven dimensional supersymmetry algebra and in the above equation we  recognise the right hand side as the result of the  the anti-commutators of two supersymmetry generators; it is the most general symmetric matrix. One can verify that the local transformations of these generators of the $l_1$ representations,  given in equation (2.24), can be written as  [8]
$$
[S_{a_1a_2a_3} , Z_\alpha{}^\beta ]
= \{{\gamma_{a_1a_2a_3}\over 2} , Z\}_\alpha{}^\beta 
\eqno(3.29)$$
This labeling of the generators as a bispinor implies a corresponding
labeling of the coordinates and so the generalised vielbein which we can
define as 
$$
{\cal V}_s=dz^\Pi E_\Pi {}^A L_A\equiv dz^\Pi E_\Pi {}_\beta{}^\alpha Z_\alpha{}^\beta ={1\over 32} Tr ( E_\Pi Z)
\eqno(3.30)$$
Comparing with the expression of equation (3.18) we find that 
$$
E_\Pi {}_\beta{}^\alpha= (\gamma_a E_\Pi {}^a- \gamma^{a_1a_2 } E_\Pi {}_
{a_1a_2 }+ \gamma^{a_1\ldots a_5 } E_\Pi {}_{a_1\ldots a_5 })_\beta{}^\alpha
\eqno(3.31)$$
we have used that 
$$
{1\over 32} Tr(\gamma^{a_1\ldots a_p}\gamma_{b_1\ldots b_p})= (-1)^{{p(p-1)\over 2}} p!\delta^{a_1\ldots a_p}_{b_1\ldots b_p}
\equiv(-1)^{{p(p-1)\over 2}} p!\delta^{a_1}_{[ b_1}\ldots \delta ^{
a_p}_{b_p]}
\eqno(3.33)$$
Using equations (3.12) and (3.29)  we find that the infinitesimal
transformation of the generalised vielbein under a local transformation
when written  in terms of the bispinor notation is given by 
$$
\delta E_\Pi {}_\beta{}^\alpha=
 {1\over 2} \{\gamma^{a_1a_2a_3}\Lambda_{a_1a_2a_3} , E_\Pi\}_\beta{}
^\alpha
\eqno(3.34)$$
We can define the inverse generalised  vielbein by $ E_\Pi
{}_\beta{}^\alpha (E^{-1})_\alpha{}^\beta{}^\Lambda= \delta ^\Lambda_\Pi$
and it transforms under a local transformation as 
$$
\delta (E^{-1})_\alpha{}^\beta{}^\Lambda=-{1\over 2} \{\gamma^{a_1a_2a_3}\Lambda_{a_1a_2a_3} , (E^{-1})^\Lambda \}_\alpha{}^\beta{}
\eqno(3.35)$$
\par
Let us  also reformulate  the transformations  of the $E_{11}$ part of
the Cartan form which is in the coset, that is the
object $P$ contained in equation (3.22), when restricted to  contain only
the three and six form fields  in terms of gamma  matrices. Let us define
$$
{\cal P}_\Pi{}_{, \alpha}{}^\beta= ({\gamma^{a_1a_2a_3}\over 2} G_{\Pi, a_1a_2a_3} 
+{\gamma^{a_1\ldots a_6}\over 4} G_{\Pi,  a_1\ldots a_6} )_{\alpha}{}^\beta
\eqno(3.36)$$
One can then verify that the transformation 
$$
\delta {\cal P}_\Pi{}_{, \alpha}{}^{\beta} = {1\over
2}[\gamma^{a_1a_2a_3}\Lambda_{a_1a_2a_3},{\cal P}_\Pi]_\alpha{}^\beta
\eqno(3.37)$$
leads to the transformations of equation (3.24) for the parts of the Cartan forms corresponding to the three form and six form fields.
\par
As discussed above we can  convert the world index on the Cartan form into a tangent space index using the inverse generalised vielbein;   using the  bispinor notation we  define 
$$
{\cal P}_\alpha{}^\beta{}_{,\gamma}{}^\delta \equiv (E^{-1})_\alpha{}^\beta {}^\Pi 
{\cal P}_\Pi{}_{, \gamma}{}^{\delta}
\eqno(3.38)$$
Let us define 
$$
{\cal P}_\alpha{}^\beta\equiv {\cal P}_\alpha{}^\delta{}_{,\delta }{}^\beta
\eqno(3.39)$$
which, using (3.35) and (3.37),  we find to   transform as
$$
\delta {\cal P}_\alpha{}^\beta = -{1\over 2} \{
\gamma^{a_1a_2a_3}\Lambda_{a_1a_2a_3},{\cal
P}\}_\alpha{}^\beta\eqno(3.40)$$ 
Thus we have found an object which is
inert under the rigid transformations and transforms covariantly under
the local transformation and as such we have found a candidate for the
equation of motion. 
 In fact we have two possible covariant objects  as we can symmetrise 
and anti-symmetrise on the $\alpha $ and $\beta $ indices after lowering
the latter with the inverse charge conjugation matrix. We note that 
the eleven dimensional gamma matrices $\gamma^{a}C^{-1}$, 
 $\gamma^{a_1a_2}C^{-1}$ and  $\gamma^{a_1\ldots a_5}C^{-1}$ are symmetric
while $C^{-1}$, $\gamma^{a_1a_2a_3}C^{-1}$ and  $\gamma^{a_1\ldots
a_4}C^{-1}$ 
 are anti-symmetric. Let us consider  the anti-symmetric part 
 which we can set to zero to obtain the invariant equation   
$$
0={1\over 2} ({\cal P}_\alpha{}^\delta (C^{-1})_\delta{}_\gamma- 
{\cal P}_\gamma{}^\delta (C^{-1})_\delta{}_\alpha )
$$
$$
= (\gamma^{a_1a_2a_3a_4}C^{-1})_\alpha{}_\gamma {\cal
P}_{a_1a_2a_3a_4}+  (\gamma^{a_1a_2a_3}C^{-1})_\alpha{}_\gamma {\cal
P}_{a_1a_2a_3}+(C^{-1})_\alpha{}_\gamma {\cal P}
\eqno(3.41)$$
Thus we find the equations 
$$
2{\cal P}_{a_1a_2a_3a_4}\equiv G_{[a_1, a_2a_3a_4 ]}
-{3.5\over 2} G_{b_1 b_2, }{}^{b_1b_2}{}_{ a_1 a_2a_3a_4 } 
-{1\over 2.4!} \epsilon_{a_1 a_2a_3a_4} {}^{b_1\ldots b_7} G_{b_1, b_2\ldots b_7 }
$$
$$-{1\over 2}  G_{b_1 b_2 [  a_1a_2a_3 , } {}^{b_1b_2}{}_{a_4 ]}
+{5 \over 4.4!} \epsilon_{a_1 a_2a_3a_4} {}^{b_1\ldots b_7} G_{c_1c_2 b_1b_2b_3 ,}{}^{c_1c_2}{}_{ b_4\ldots b_7}=0
\eqno(3.42)$$
$$
2{\cal P}_{a_1a_2a_3}\equiv -6G_{[a_1|b|, }{}^{b}{}_{a_2a_3]}
+{1\over 4}\epsilon_{a_1 a_2a_3} {}^{b_1\ldots b_8} G_{b_1 b_2. b_3\ldots b_8 }
$$
$$
+{1\over 3!.5!} \epsilon_{a_1 a_2a_3} {}^{b_1\ldots b_8} G_{b_1 \ldots b_5 ,b_6\ldots b_8 }
+{5.3\over 2} G^{c_1c_2c_3c_4}{}_{[a_1, |c_1c_2c_3c_4|a_2a_3 ]}=0
\eqno(3.43)$$
and 
$$
2{\cal P}\equiv {1\over 2.5!.11!}\epsilon^{a_1\ldots a_{11}} G_{a_1\ldots a_5, a_6\ldots a_{11}}=0
\eqno(3.44)$$
In finding these equations use was made of the identity
$$
\gamma^{a_1\ldots a_p} \gamma_{b_1\ldots b_q}
=\sum_r (-1)^{{r(r-1)\over 2}}(-1)^{(p-r)r}{p!q!\over r!(p-r)!(q-r)!}
\delta^{[a_1\ldots a_r}_{[b_1\ldots b_r}\gamma^{a_{r+1}\ldots a_p ]}
{}_{b_{r+1}\ldots a_q ]}
\eqno(3.45)$$
and equation (3.31). In considering these equations it is important to
recall that we have set to zero all contributions involving the gravity
and dual gravity fields. 
\par
We note that only the first of these equations involves the derivative with respect to the usual coordinates $x^a$ of space-time. At the linearised level this equation  is given by 
$$
\partial_{[a_1} A_{a_2a_3a_4 ]}
-{3.5\over 2} \partial_{b_1b_2} A^{b_1b_2}{}_{ a_1, a_2a_3a_4 } 
-{1\over 2.4!} \epsilon_{a_1 a_2a_3a_4} {}^{b_1\ldots b_7}
 \partial_{[b_1}A_ {b_2\ldots b_7 ]}
$$
$$
-{1\over 2}  \partial_{b_1b_2 [  a_1a_2a_3  } A^{b_1b_2}{}_{a_4 ]}
+{5\over 4.4!}  \epsilon_{a_1 a_2a_3a_4} {}^{b_1\ldots b_7}
\partial_{c_1c_2 b_1b_2b_3 }A^{c_1c_2}{}_{ b_4\ldots b_7}=0 
\eqno(3.46)$$
where $\partial_a={\partial\over \partial x^a}$, 
$\partial^{ab}={\partial\over \partial x_{ab}}$ and $\partial^{a_1\ldots
a_5}={\partial\over \partial x_{a_1\ldots a_5}}$.  
\par
If we were to restrict the dependence of the fields to only be on $x^a$ then the last equation would be the correct equation of motion for the three and six form fields at linearised level, Indeed at the full non-linear level we find the field equation 
$$
 F_{c_1\ldots c_4}={1\over 7.6.2}
\epsilon _{c_1\ldots c_4}{}^{b_1\ldots b_7} F_{b_1\ldots b_7}
\eqno(3.47)$$
where 
$$ F_{c_1\ldots c_4}
\equiv  4 \partial_{ [ c_1} A_{c_2\ldots c_4]}
\eqno(3.48)$$
and 
$$
 F_{c_1\ldots c_7}\equiv  
7(\partial_{[ c_1} A_{c_2\ldots c_7]}-A_{[ c_1c_2c_3}\partial_{c_4}A_{c_5c_6c_7 ]})
\eqno(3.49)$$
\par
The reader may wish to explicitly vary equation (3.42) using equations
(3.24) and (3.26) and  show that the resulting terms which  contain  the
usual space-time derivative $\partial_a$ do actually cancel. A useful
intermediate result is  that 
$$
\delta (G_{[a_1, a_2a_3a_4 ]}-{3.5\over 2} G_{b_1b_2, }{}^{b_1b_2}{}_{ a_1, a_2a_3a_4 } )= +{5!.7\over 8}G_{[b_1, b_2b_3 a_1\ldots a_4]}\lambda ^{b_1 b_2b_3}
\eqno(3.50)$$
from which we see that although the left-hand side is not a field
strength both terms conspire so as to give a variation that is a field
strength,  as the equation of motion requires. 
\par
The most general invariant equation linear in generalised space-time
derivatives would be a sum of the symmetric and antisymmetric parts of 
${\cal P}_\gamma{}^\delta (C^{-1})_\delta{}_\alpha$ with arbitrary
coefficients.  However, the symmetric part involves 
 terms such as 
$G_{b, a_1a_2a_3} \eta^{ba_1}$ which are clearly not gauge invariant  if
one restricts the dependence on the generalised space-time to be only  on
$x^a$. The strategy used here is similar to that used in the original
paper [14] on gravity, except that they used conformal symmetry to fix
the constants, and the approach used in the early $E_{11}$ papers, such
as  [1,12]. However, an extension of this procedure was applied in [15,16]
where the non-linear realisation of $E_{11}\otimes_s l_1$,  appropriate to
four dimensions and at lowest level, was carried out. This meant keeping
only  the coordinates of the four dimensional space-time and those in the
56 dimensional representation of $E_7$  from the $l_1$ representations. 
An invariant action was then found that contained a number of
undetermined constants. The constants were then fixed by demanding
general coordinate invariance once the fields had been restricted to
depend on the coordinates of the four dimensional space-time and only the
usual seven of the 56 other coordinates. It is this strategy we are
following here. 
\medskip
{\bf 4. Discussion }
\medskip 
The variation of equation (3.42) under the transformations of equation
(3.24) will lead to a duality relation between  the derivative of the
graviton and that of the dual graviton. It would certainly be interesting
to  find what these equations are and if they really do describe the
correct dynamics for gravity as it appears in the framework of eleven
dimensional supergravity once we neglect the higher $E_{11}$ fields and
the higher coordinates.  Should this be the case then the $E_{11}$
conjecture [1,8] will be shown to be true, namely that the non-linear
realisation of $E_{11}\otimes_s l_1$ is an extension of the dynamics of
eleven dimensional supergravity.    We hope to report on this
soon. Although when $E_{11}$ was first proposed the meaning of the higher
fields was unknown,  we have now come to understand the physical
significance for large numbers of the higher level fields. The result in
this paper suggests that the additional coordinates will also have a
physical meaning.  It would be very straightforward to extend the 
results in this paper  to the IIA and IIB theories in ten dimensions and
the theories in $d$ dimensions using the techniques previously employed
[1,2,3,4,5,6,7]. 
\par
The non-linear realisation of  $E_{11}\otimes_s l_1$  has not been
systematically computed before. In the early papers on $E_{11}$ only the
coordinate $x^a$ was used and the local subalgebra was taken to be just
the Lorentz algebra. As a result much of the power of the non-linear
realisation was lost. Nonetheless many of the features of the
supergravity theories were recovered, for example the fields strengths
for all  gauged supergravity theories in five dimensions [5]. This paper
should open the way to the systematic computation of the $E_{11}\otimes_s
l_1$  non-linear realisation and so the dynamics it contains. 
\par
Above we simply deleted the dependence of  the fields on the  higher
coordinates. However, it remains to understand what physical procedure
one should use to reduce the  dependence on the fields on the generalised
space-time.  The  work of reference [17] suggested that even thoughÊ the
full theory was
$E_{11}\otimes_s l_1$ invariant only part of the $l_1$ representation
occurred in the second quantised field theory. In particular although
the first quantised theory involved all of the coordinates of the $l_1$
representation, the choice of representation of the commutators that
takes one  from the
first to the second quantised theory required one to choose only part of
the $l_1$ representation. However, one canÊ make different choices of
which part of the $l_1$ representation one takes and these should be
equivalent and related by
$E_{11}$ transformations. It would be interesting to  really understand
how this works. However, it is likely that  a simple truncation will not
be the only allowed possibility; indeed  in  the
construction of  all the five dimensional supergravity theories [5]
we found a much more subtle procedure involving a slice that included
part of $E_{11}$. 
\par
To close it could be helpful  to discuss the relationship between the
various works on generalised geometry. This paper is based on the 2003
proposal to consider the non-linear realisation of $E_{11}\otimes_s l_1$
[8], however, there are several other approaches. A generalised
space-time appeared in the context of   string dynamics where the usual
space-time was extended to include  an additional coordinate $y_a$. This
was done  in such a way as to encode the (first quantised) dynamics of
the string in an O(D,D) symmetric manner [18,19];  a generalisation to
the membrane was also given [20].  In fact the dynamics of strings and 
membranes can be formulated as a non-linear realisation of
$E_{11}\otimes_s l_1$ [21], The difference from  the non-linear
realisation studied above is that a different 
 choice of local subalgebra is taken and  the coordinates 
associated with the
$l_1$ representation become fields.  The non-linear realisation can be
carried out so as to include the background supergravity fields that
belong to the $E_{11}$ part of the non-linear realisation in the same 
way as above.  If one takes the non-linear realisation of $E_{11}\otimes_s
l_1$ appropriate to ten dimensions and at lowest level one finds the
generalised space-time and the string dynamics given in [18,19,20].
Carrying out this non-linear realisation  in $d$ dimensions at lowest
order one finds the coordinates of
$d$ dimensional Minkowski space-time and in addition scalar coordinates 
belong to the 10, $\bar {16}$,  $\bar {27}$, 56 and $248\oplus 1$ of
SL(5), SO(5,5),
$E_6$. $E_7$ and $E_8$ for $d$ equal to  seven, six, five and four and
three dimensions respectively [3,11,21,5]. These are the same coordinates
as arises at level zero in  the non-linear realisation of $E_{11}\otimes_s
l_1$  used to find the supergravity theories in $d$ dimensions in the
absence of strings and branes. 
\par
Another version of generalised geometry was  inspired by a version of
closed string field theory [22] and the papers of [23]. It goes by the
name of doubled field theory as it also doubles the number of
coordinates to have a $x^a$ and $y_a$ in order to encode an  O(10,10)
symmetry [24]. The   field theory is defined on this space contains 
the same fields as in  the NS-NS sector of the ten dimensional
superstring. An action was constructed and  if one restricts the fields to
depend on just  the usual space-time, that is  just on  $x^a$,  one
finds  the well known action for the NS-NS sector [33,34]. However,
doubled field theory  is just a sub case of the non-linear realisation of
$E_{11}\otimes _sl_1$. To be precise it is the non-linear realisation of
$E_{11}\otimes _sl_1$ at lowest level in the decomposition appropriate to
the  IIA theory [25]. This is a very straightforward systematic 
calculation that requires no guess work and took only six pages in [25] to
present in all detail. The advantage of viewing this as a non-linear
realisation of
$E_{11}\otimes _sl_1$ is that it is places  the construction in a wider
conceptual framework in which the true nature of the symmetries is
apparent. For example, the presence of the  GL(1) symmetry in
addition to O(10,10) becomes clear and one can construct
the  extension to the next level [26]. This is also very straightforward
and  one finds [26] the R-R part of the well known supergravity equations
of motion.  One could  also compute even higher levels involving fields
beyond that of the usual maximal supergravity theories and so find new
physics. 
\par
The approach   of Ê[19,20] just mentioned above   had the aim of encoding
some duality symmetries in the first quantised dynamics. This work was
taken up in [27] which derived a generalised metric from the first
quantised theory and used this to construct invariant Sl(5) dynamics for
fields living on a space whose coordinates belonged to Êthe Êten
dimensional representation of SL(5). 
 This work was then generalised to the duality group SO(5,5)
[28]. This was in agreement with the general framework of [21] and the
non-linear realisation of $E_{11}\otimes _sl_1$ taking into account
that the coordinates of the $l_1$ representations in $d$
dimensions at level zero are those    mentioned just above. Very
recently  it was shown [29] in detail how these theories [27,28] were
the non-linear realisation of
$E_{11}\otimes _sl_1$ appropriate to seven and six dimensions  at lowest
level. The work of [29] also contained the   generalisation  to find the
analogues of these results in five and four dimensions and so involving
the duality groups $E_6$ and $E_7$ respectively. The precise relationship
to the work of [15,16] which earlier  computed the non-linear realisation
of
$E_{11}\otimes _sl_1$ at lowest level in four dimensions and used the
generalised space consisting of the usual coordinates of space-time and
the coordinates in the 56 dimensional representation of
$E_7$ has yet to be clarified. 

There is yet another approach  inspired by the work of 
Hitchin [30] and Gualtieri [31]. This introduced an extended tangent
space, associated with O(D,D) but does  not extend our usual notion of
space-time, see for example  [32] and references therein.  There has
not been a study to investigate the connection to  the non-linear
realisation of $E_{11}\otimes_s l_1$. 
However,   the generalised   tangent
space in ten dimensions  which encodes $O(D,D)\otimes GL(1)$ with
tangent group $O(10)\otimes O(10)$ [32], is precisely the  same  as that
which arises in the  non-linear realisation of
$E_{11}\otimes_s l_1$ appropriate to ten dimensions at lowest level 
[25,26]. As we mentioned below equation (3.19) the generalised
tangent space   is in general  just that given by the $l_1$
representation and the tangent space group is $I_c(E_{11})$. In eleven
dimensions this is just the usual  tangent space,  the space of two forms
and five forms and higher objects [8]. While in $d$ dimensions we would
find the usual tangent space,   scalars   belong to the 10, $\bar {16}$, 
$\bar {27}$, 56 and
$248\oplus 1$ of SL(5), SO(5,5),
$E_6$. $E_7$ and $E_8$ for $d$ equal to seven, six, five,  four and
three dimensions respectively as well as higher level objects 
[3,21,5,11].  This leads one to suspect that if one carries out 
the $E_{11}\otimes_s l_1$ non-linear realisation, but at the end sets all
the fields to depend  on just  the usual coordinates $x^a$ then one
might obtain this approach. 

\medskip
{\bf Acknowledgments}
\medskip
 I wish to thank Dario Martelli for discussions and  STFC for support 
from the grant given to the theoretical physics group in the Mathematics
Department at King's. 

\medskip
{\bf References}
\medskip
\item{[1]} P. West, {\it $E_{11}$ and M Theory}, Class. Quant.
Grav. {\bf 18 } (2001) 4443,  hep-th/010408.
\item{[2]} I. Schnakenburg and P. West,{\it ÊKac-Moody Symmetries of
IIB Supergravity}, ÊPhys.Lett. {\bf 517B} (2001) 421-428, hep-th/
0107181.
\item{[3]} P. West,  {\it $E_{11}$ origin of Brane charges and U-duality
multiplets}, JHEP 0408 (2004) 052, hep-th/0406150. 
\item{[4]}  F. ÊRiccioni and P. West, {\it The $E_{11}$ origin of all
maximal supergravities}, ÊJHEP {\bf 0707} (2007) 063; ÊarXiv:0705.0752.
\item{[5]} ÊF. Riccioni and P. West, {\it E(11)-extended spacetime
and gauged supergravities},
JHEP {\bf 0802} (2008) 039, ÊarXiv:0712.1795
\item{[6]} ÊF. Riccioni and P. West,
Ê{\it Local E(11)}, JHEP {\bf 0904} (2009) 051, arXiv:hep-th/0902.4678.
\item{[7]} F. ÊRiccioni, ÊD. ÊSteele and P. West, {\it The E(11)
origin of all maximal supergravities - the hierarchy of field-strengths}
ÊÊJHEP {\bf 0909} (2009) 095, arXiv:0906.1177. 
\item{[8]} P. West, {\it $E_{11}$, SL(32) and Central Charges},
Phys. Lett. {\bf B 575} (2003) 333-342, {\tt hep-th/0307098}
\item{[9]}  A. Kleinschmidt and P. West, {\it  Representations of G+++ and
the role of space-time},  JHEP 0402 (2004) 033,  hep-th/0312247.
\item {[10]} P. West, {\ it The IIA, IIB and eleven dimensional theories 
and their common
$E_{11}$ origin}, Nucl. Phys. B693 (2004) 76-102, hep-th/0402140. 
\item{[11]} P. Cook and P. West, {\it Charge multiplets and masses
for E(11)}, ÊJHEP {\bf 11} (2008) 091, arXiv:0805.4451.
\item{[12]} P.~C. West, {\sl Hidden superconformal symmetry in {M}
    theory },  JHEP {\bf 08} (2000) 007, {\tt hep-th/0005270}
\item {[13]} P. West, {\it Very Extended $E_8$ and $A_8$ at low
levels, Gravity and Supergravity}, Class.Quant.Grav. {\bf 20} (2003)
2393, hep-th/0212291.
\item{[14]} V. Ogievetsky, Lett. {\it Infinite-dimensional algebra of
general  covariance group as the closure of the finite dimensional
algebras  of conformal and linear groups}, Nuovo. Cimento, 8 (1973) 988; 
 A. Borisov and V. Ogievetsky,  {\it Theory of dynamical affine and
conformal  symmetries as the theory of the gravitational field}, 
Teor. Mat. Fiz. 21 (1974) 32
\item {[15]} C. Hillmann, {\it Generalized E(7(7)) coset dynamics and D=11
supergravity}, JHEP {\bf 0903}, 135 (2009), hep-th/0901.1581. 
\item{[16]} C. Hillmann, {\it E(7(7)) and d=11 supergravity }, PhD
thesis,  arXiv:0902.1509. 
\item {[17]} P. West, {\it Generalised space-time and duality},
hep-th/1006.0893.  
\item{[18]} A. Tseytlin, {\it Duality Symmetric Formulation Of String World
Sheet Dynamics}, Phys.Lett. {\bf B242} (1990) 163, {\it Duality Symmetric
Closed String Theory And Interacting Chiral Scalars}, Nucl.\ Phys.\ B {\bf
350} (1991) 395. 
\item{[19]} M. Duff, {\it Duality Rotations In String Theory},
  Nucl.\ Phys.\  B {\bf 335} (1990) 610; 
\item{[20]} M. Duff and J. Lu,
 {\it Duality rotations in
membrane theory},  Nucl. Phys. {\bf B347} (1990) 394. 
\item{[21]} P. West, {\it Brane dynamics, central charges and
$E_{11}$}, hep-th/0412336. 
\item{[22]} T. Kugo and B. Zwiebach,{\it Target space duality as a symmetry
of string field theory},
  Prog.\ Theor.\ Phys.\  {\bf 87}, 801 (1992) hep-th/9201040.
\item{[23]}  W. Siegel, {\it Superspace duality in low-energy superstrings},
  Phys.\ Rev.\  D {\bf 48}, 2826 (1993), hep-th/9305073;
 {\it  Two vierbein formalism for string inspired axionic gravity},
  Phys.\ Rev.\  D {\bf 47}, 5453 (1993) hep-th/9302036.
\item{[24]} C. Hull and B. Zwiebach, {\it Double Field Theory},  JHEP {\bf 0909}
(2009) 099, hep-th/0904.4664;   C. Hull and B. Zwiebach, {\it The gauge algebra of double field
theory and Courant brackets},
  JHEP {\bf 0909} (2009) 090, hep-th0908.1792; 
 O. Hohm, C. Hull and B. Zwiebach, {\it Background independent
action for double field theory}, hep-th/1003.5027; 
 O. Hohm, C. Hull and B. Zwiebach, {\it Generalised metric
formulation of double field theory},  hep-th/1006.4823; 
\item{[25]}  P. West, {\it E11, generalised space-time and IIA string theory}, 
 Phys.Lett.B696 (2011) 403-409,   arXiv:1009.2624.
\item{[26]}   A. Rocen and P. West,  {\it E11, generalised space-time and IIA string theory; 
the R-R sector},  arXiv:1012.2744.
\item{[27]}
D.~S.~Berman, M. Perry, {\it Generalized Geometry and M theory}, 
ÊÊJHEP {\bf 1106} (2011) 74
ÊÊ[arXiv:1008.1763 [hep-th]].
\item{[28]}
D.~S.Berman, H. Godazgar and M. Perry,
{\it SO(5,5) duality in M-theory and generalized geometry}, 
ÊÊPhys.\ Lett.\ Ê{\bf B700 } (2011) Ê65-67.
ÊÊ[arXiv:1103.5733 [hep-th]].
\item{[29]} D.  Berman, H. Godazgar, M. Perry and  P.  West, 
{\it Duality Invariant Actions and Generalised Geometry}, 
arXiv:1111.0459. 
 \item{[30]} N. Hitchin, Generalized Calabi-Yau manifolds, 
 Q. J. Math.  {\bf 54}  (2003), no. 3, 281,
math.DG/0209099;  {\it Brackets, form and
invariant functionals}, math.DG/0508618.
\item{[31]} M. Gualtieri, {\it Generalized complex geometry}, PhD Thesis
(2004), math.DG/0401221v1. 
\item{[32]} A. Coimbra, C. Strickland-Constable and  D.  Waldram, 
{\it Supergravity as Generalised Geometry I: Type II Theories}, 
arXiv:1107.1733.
\item{[33]} J. Schwarz and P. West,
  {\it Symmetries And Transformations Of Chiral N=2 D = 10 Supergravity}, 
  Phys.\ Lett.\ B {\bf 126} (1983) 301; 
P. Howe and P. West,
  {\it The Complete N=2, D = 10 Supergravity}, 
  Nucl.\ Phys.\ B {\bf 238} (1984) 181;
 J. Schwarz,
  {\it Covariant Field Equations Of Chiral N=2 D = 10 Supergravity}, 
  Nucl.\ Phys.\ B {\bf 226} (1983) 269.
 \item{[34]}
  I. Campbell and P. West,
  {\it N=2 D = 10 Nonchiral Supergravity And Its Spontaneous
Compactification}, 
  Nucl.\ Phys.\ B {\bf 243} (1984) 112;
 F. Giani and M. Pernici,
  {\it N=2 Supergravity In Ten-Dimensions}, 
  Phys.\ Rev.\ D {\bf 30} (1984) 325; M. Huq and M. Namazie,
  {\it Kaluza-Klein Supergravity In Ten-Dimensions}, 
  Class.\ Quant.\ Grav.\  {\bf 2} (1985) 293.
  
 \eject 
{\bf N. Note added  on the equation of motion in the gravity sector}
\medskip

In this note added we will derive,  from the $E_{11}\otimes_s l_1$
non-linear realisation, the equation of motion that relates the
usual field of gravity to its dual field.  We will work
only to the level that the dual graviton occurs and so  we will not derive
the terms involving  the higher level
$E_{11}$ fields associated with gravity,  nor shall we derive the terms
that contain generalised space-time derivatives for coordinates beyond
the lowest level, that is, we will only find terms with derivatives with
respect to the coordinates of our customary  space-time. 

The notion of a dual graviton was first introduced by Curtwright [N1].
This work was based on an equation that involved the Riemann tensor. The 
dual graviton
$h_{a_1\ldots a_8, b}$ arises   automatically as a field at level three in
the $E_{11}$ non-linear realisation and  it was proposed that it was
related to the vielbein by a duality relation that  is  first order in
derivatives  [1]. 
 In reference [1] a   non-linear  equation, involving the vielbein and a
field $Y_{b_1\ldots b_9, b}$, which did correctly  described full
gravity   was given and  it was also shown  that  at the
linearised level could one substitute the field $Y_{b_1\ldots b_9, b}$
for the derivative of the dual gravity field and find the correct
equation for linearised gravity.  This left  the
situation for the full non-linear theory unresolved. However, in
reference [13] an equation involving the vielbein, its dual and the field
$Y_{b_1\ldots b_9, b}$, which was first order in derivatives,  was given
and shown the correctly describe gravity at all orders. This result seems
to have gone  somewhat unnoticed in the subsequent literature. 
\par
Although these results were encouraging,  there remained the problem of
how a duality  relation between the vielbein and the dual gravity field
would arise in
$E_{11}$. This was discussed in reference   [N2] and in [N3]; the latter 
includes considerations based on supersymmetry, but no firm contact with
$E_{11}$ was found.

Equation (3.42),  which relates the three form gauge field to its dual
six form field,   was derived by introducing objects with  spinor indices
in a maneuver that may not generalise to the other fields. As such we
begin this section by giving a more conventional derivation of this
equation of motion. In particular we will explicitly vary equation (3.42)
under the local $I_c(E_{11})$ symmetry  variations of equations (3.24) and
(3.26). We recall that the way the dynamical equations are constructed
ensures that they are automatically invariant under the rigid
$E_{11}\otimes_s l_1$ transformations. The equation is built from the
Cartan forms of equation (3.14-16) whose indices are of the form
$G_{\diamond ,\bullet}$. The $\bullet $ represent the indices that are
contracted with the $E_{11}$ generators  that occur in the
Cartan form and their variations under
$I_c(E_{11})$ are given in equation (3.24). The $\diamond  $ indices
arise from the forms,  when converted to tangent indices with the inverse
generalised vielbein,  that occur in the Cartan form. 
These latter indices transform under the local transformations  given in
the second line of equation (3.26). To give an example,  
$G_{a_1a_2, b_1b_2b_3}$ occurs with  the $E_{11}$ generators
$R^{b_1b_2b_3}$ and the two form $dx^\pi E_\pi {}^{a_1a_2}$ in the Cartan
form. For
simplicity we will often denote a generic Cartan form  by just indicating
the number of indices, for example we might denote the object just
discussed by  
$G_{2,3}$. We note that this Cartan form contains generalised  space-time
derivatives that are with respect to
$x^{a_1a_2}$ and higher level coordinates, that is it does not contain
derivatives with respect to the  coordinates $x^a$ of the
usual space-time. Similar statements apply to all the  other Cartan
forms. Of course in principle one should write the variations of the
$\bullet$ and
$\diamond$ indices   in  one equation, but it can be convenient to carry
them out separately.  
\par
We now carry out the variation of the equation (3.42) under the local
$I_c(E_{11})$ transformations for both  of the two  types of
variation just mentioned, but keeping terms in the variation that contain
Cartan forms  $G_{\diamond , \bullet}$ with a $\diamond$ index  that is
an $a$ or an
$a_1a_2$ , that is,  we discard Cartan forms  associated with
generalised space-time derivatives with respect to
$x_{a_1\ldots a_5}$ and higher level coordinates. We will first focus on
terms in the variation that contain the three and six form gauge fields
and later collect the terms that contain the gravity field and its dual. 
The  terms in the variation that contain
$G_{2,3}\Lambda_3$ are given by 
$$
+9G_{[a_1a_2, }{}^{ b_1b_2 }{}_{a_3}\Lambda _{a_4 ] b_1b_2}
+9G^{b_1b_2}{}_{, [a_1a_2|b_1}\Lambda _{b_2 |a_3a_4]}
-18 G_{[a_1| b,}{}^{bc}{}_{a_2}\Lambda _{|c| a_3a_4]}
\eqno(N.1)$$
In deriving this result we have already implemented  the
cancellations that occur between certain terms. While  the terms involving
$G_{2,6}\Lambda_3$ are given by 
$$
+{1\over 4.4} \epsilon _{a_1a_2a_3a_4}{}^{b_1\ldots b_7} \{
\Lambda_{ b_1c_1c_2}G^{c_1c_2}{}_{,b_2\ldots b_7}
+15\Lambda_{ b_1c_1c_2}G_{b_2b_3}{}^{c_1c_2}{}_{,b_4\ldots b_7}
+5.6 \Lambda _{c_2b_1b_2} G_{b_3c_1, }{}^{c_1c_2}{}_{b_4\ldots b_7} \}
\eqno(N.2)$$
where again we have not shown terms that cancel against each other. 
\par
The second and third terms of equation (N.1)  can be written as 
$$
-9 G_{[a_1|b, }{}^b{}_{|c a_2] }\Lambda ^{c}{}_{a_3a_4}
\eqno(N.3)$$
where the anti-symmetry on $a_1, a_2, a_3$ and $a_4$ in this particular
equation is not indicated explicitly but is to be understood to be
present.

The first and second terms in equation (N.2) can be written as 
$$
+{7\over 4} \epsilon _{a_1a_2a_3a_4}{}^{b_1\ldots b_7} \{
\Lambda_{ b_1}{}^{c_1c_2}G_{[ c_1c_2,b_2\ldots b_7]}
-{6.4\over 8.7} \Lambda_{ b_1c_1c_2}G_{b_2 c_1, b_3\ldots b_7}\}
\eqno(N.4)$$
To lowest order equation (3.43) can be expressed as 
$$
G_{[b_1b_2, b_3\ldots b_8]}=-{6.4\over 8!. 3!} \epsilon _{b_1\ldots
b_8}{}^{e_1e_2e_3} G_{e_1b,} {}^b{}_{e_2e_3}
\eqno(N.5)$$
and substituting this into the  first term in equation (N.4) we find that
it  cancels the  terms two and three in equation (N.1).  
\par
The net effect of all this is that the local $I_c(E_{11})$ variation of
equation (3.42) which contain only terms that involve the three and six
form gauge fields, that is the Cartan  forms
$G_{2,3}$ and
$G_{2,6}$,  is given by   
$$
+9G_{[a_1a_2, }{}^{ b_1b_2 }{}_{a_3}\Lambda _{a_4 ] b_1b_2}
+{6\over 4.4} \epsilon _{a_1a_2a_3a_4}{}^{b_1\ldots b_7}\{ 
-2\Lambda_{ b_1c_1c_2}G_{b_2 c_1, b_3\ldots b_7}
+5\Lambda _{b_1b_2 d} G_{b_3c, }{}^{c d}{}_{b_4\ldots b_7}
\}
\eqno(N.6)$$
These are  the first term in equation (N.1),  the second term in
equation (N.4) and the third term in equation (N.2). 
These terms can be
canceled by introducing gravity and dual gravity terms into equation
(3.42) whose Cartan forms have the local variations of equations (3.24)
and (3.26). Carrying out these modifications 
equation (3.42) is now  given by  
$$
0= G_{[a_1, a_2a_3a_4 ]}
-{3.5\over 2} G_{b_1 b_2, }{}^{b_1b_2}{}_{ a_1 a_2a_3a_4 } 
-{1\over 2.4!} \epsilon_{a_1 a_2a_3a_4} {}^{b_1\ldots b_7} G_{b_1, b_2\ldots b_7 }
$$
$$-{1\over 2}  G_{b_1 b_2 [  a_1a_2a_3 , } {}^{b_1b_2}{}_{a_4 ]}
+{5 \over 4.4!} \epsilon_{a_1 a_2a_3a_4} {}^{b_1\ldots b_7} 
G_{c_1c_2 b_1b_2b_3 ,}{}^{c_1c_2}{}_{ b_4\ldots b_7}
$$
$$
-{1\over 2} G_{[a_1a_2,a_3a_4]}
-{7\over 6} \epsilon _{a_1a_2a_3a_4}{}^{b_1\ldots b_7}
G_{c b_1, }{}^c{}_{b_2\ldots b_7 d, }{}^d
\eqno(N.7)$$
The variation of the second to last  and last terms in this
equation cancel the first and second variations of equation (N.6)
respectively. Thus we have found an equation for the three form and six
form gauge fields that is invariant under the local $I_c(E_{11}) $
transformations and so all the transformations of the non-linear
realisation if we discard in the variation terms containing gravity 
and its dual and generalised space-time derivatives with respect to 
coordinates beyond the two form. 
\par
To better understand the above calculation it is useful to represent it
in  a schematic diagram, which is  given in equation (N.8). We searching
for an equation with four anti-symmetric indices constructed from the
Cartan forms
$G_{\diamond,
\bullet}$ and the epsilon symbol, denoted by 
$\star$. In the diagram below we list all possible terms in a grid
going with increasing level of the fields to the right and increasing
level in the generalised space-time derivatives as one goes down. Where
there is no term indicated it means that there is no such term that one
can write down with the correct indices. 
$$
\matrix {&\leftarrow &G_{1,3}&\leftrightarrow&\star G_{1,6}
&\rightarrow \cr 
\uparrow& &\downarrow& & \updownarrow & &\uparrow\cr
G_{2,1,1}&\rightarrow& & \leftarrow& G_{2,6}& \leftrightarrow &
\star G_{2,8,1}\cr
\updownarrow& &\uparrow& & \updownarrow & &\updownarrow\cr
\star G_{5,1,1}&\leftrightarrow &G_{5,3} & \leftrightarrow& \star G_{5,6}&
\leftrightarrow & G_{5,8,1}\cr}
\eqno(N.8)$$
The arrows indicate what happens when one varies the terms under the
local transformations. In particular, the vertical arrows indicate the
effect of varying the first indices, that is the $\diamond$ indices and
the horizontal arrows the effect of varying the second indices,
that is the $\bullet$ indices. The terms so obtained contain the
Cartan field at the site the arrow point to times the parameter
$\Lambda_3$. For example, the variation of the first  index on $G_{1,3}$
leads to a term $\Lambda_3 G_{2,3}$ whose only other sources are given by
following the arrows pointing to this site; 
for example,  one such term arises from the variation of the second index
of
$G_{2,6}$.   We note that just because a Cartan form is absent in the
four index equation of motion this does not mean that the same Cartan
form does not arise in the local variation of this equation as the
variation has in general a different index structure. 
\par
Finally, we now compute the local variation of equation (3.42) keeping the
remaining terms, that is, those that contain the gravity field or its
dual. We will  
only keep derivatives with respect to the usual coordinates of space-time,
that is,  those with respect to
$x^a$. We note that the resulting equation will have  a different index
structure to that for the three form field and as a result the spaces in
the diagram  where no such contribution can exist are different. 
Varying equation (N.7) we are interested in the 
 terms correspond to the spaces in
the diagram of equation (N.8) that are in the top line at the extreme left
and right hand ends and these   can only come from the variation of
$G_{a, b,c}$ ($G_{2,1,1}$), $G_{a, b_1\ldots b_8,c}$ ( $G_{2,8,1}$), by
varying their first index, or
$G_{1,3}$ and
$G_{1,6}$ by varying their second index.  Since there are no terms in the
four index equation that involve the Cartan forms $G_{a, b,c}$  and $G_{a,
b_1\ldots b_8,c}$ the terms in the variation   must either cancel or
result in a new equation. In fact this new  equation
has three indices. The most general
terms one can write down with three indices, with no particular symmetry,
are given in the diagram of equation (N.9) whose interpretation is
analogous to that for the diagram of equation (N.8). 
$$
\matrix {G_{1,1,1}&\rightarrow & & & 
&\leftarrow &\star G_{1,8,1}\cr 
\downarrow& &\uparrow& & \uparrow & &\downarrow\cr
&\rightarrow& G_{2,3}& \leftarrow&\star G_{2,6}& \leftrightarrow &\cr
\uparrow& &\updownarrow& & \updownarrow & &\uparrow\cr
 G_{5,1,1}&\leftrightarrow &\star G_{5,3} & \leftrightarrow& 
G_{5,6}&
\leftrightarrow & \star G_{5,8,1}\cr}
\eqno(N.9)$$  
Setting the  variation  to zero we
find that  
$$
3X_{[a_1a_2,|c|} \Lambda^c{}_{a_3a_4 ]}
+{7} \epsilon _{a_1a_2a_3a_4}{}^{b_1\ldots b_7}
\{G_{b_1, c_1c_2c_3 b_2\ldots b_6,  b_7}\Lambda ^{c_1c_2c_3}
-G_{c_1, c_2 b_2\ldots b_7 d,}{}^d\Lambda^{c_1c_2 b_1}\}=0
\eqno(N.10)$$
where 
$$
X_{a_1a_2,c}= -G_{[a_1, a_2 ]c}- G_{[a_1, |c| a_2]} +G_{c, [a_1 a_2]}
\eqno(N.11)$$
We recall that $G_{c, a}{}^b= e_c{}^\mu (e\partial _\mu e)_a{}^b$
whereupon we recognise that $X_{a b,c}= w_{c, a b}$  where $ w_{c,
a}{}^b$ is the usual spin connection.  Extracting the
parameter in equation (N.10) we find the equation 
$$
3X_{[a_1a_2,}{}^{ [c_1} \delta ^{c_2c_3]}_{a_3a_4 ]}
+7 \epsilon _{a_1a_2a_3a_4}{}^{b_1\ldots b_7}
\{G_{b_1, c_1c_2c_3 b_2\ldots b_6,  b_7}
-\delta _{b_1,[c_3}G_{c_1, c_2] b_2\ldots b_7 d,}{}^d\}=0
\eqno(N.12)$$
Setting $a_3=c_2$ and $a_4=c_3$ and summing over these indices we find
that 
$$
{7.2\over 3}w^{c_1}{}_{, a_1a_2}{}-{8\over 3} w_{d [
a_1,}{}^{d}\delta_{a_2]}^{c_1}
 + \epsilon _{a_1a_2}{}^{b_1\ldots b_9}
G_{b_1,  b_2\ldots b_9,  c_1}
+{7\over 3} \epsilon _{a_1a_2}{}^{c_1}{}^{b_1\ldots b_8}
G_{b_1, b_2\ldots b_8 d,}{}^d =0
\eqno(N.13)$$
where we have substituted  the spin connection. 
\par
We see that the equation has many of the correct features and in
particular it  relates  the derivative of the vielbein, specifically the
spin connection,  to the derivative of the dual graviton field. Of course
the full equation will contain  higher level
$E_{11}$  fields and also  derivatives with respect to the higher level
coordinates of the generalised space-time.  We observe that equation
(N.13) contains not only the $P$ part of the Cartan forms of equation
(3.22) but also those of equation (3.23), that is, the $Q$ part,  which
transform inhomogeously under the local symmetry. In particular it
contains the Cartan form 
$G_{c, [ab]}$ which transforms inhomogeneously under the local Lorentz
group, see equation (3.21). The appearance of this Cartan form in the
equations of motion  is related to the fact that we have not chosen the
local Lorentz transformation to set the anti-symmetric part of the
graviton to zero, or equivalently, one of the off diagonal parts of the
vielbein to zero. It follows that  equation (N.13) does  not strictly
speaking hold as an equality as while the right-hand side transforms
covariantly under local Lorentz transformations the left-hand side
transforms  like the spin connection and so has an inhomogeneous
term of the form $\partial^{c_1}\Lambda_{a_1a_2}$ in its local variation. 
The
$E_{11}$ algebra implies that the dual gravity field satisfies the
constraint
$h_{[a_1\ldots a_8, b]}=0$, but it is easy to see that  the effect of
allowing equality up to the above term is   equivalent to relaxing this
constraint and introducing a nine form object into the theory [13].  
\par
To find an  invariant equation  one must form the Riemann tensor
$R_{\mu\nu, a}{}^b$ from the spin connection in the usual way. The
corresponding terms   involving the dual
graviton do not vanish. However they do vanish at least at the
linearised level if one then forms the usual contraction to form the
Ricci tensor,  by setting
$\nu=b$ and then summing. Thus at the
linearised level one would have the correct equation for gravity if
it were it not for the following observation.  We note that if we set 
$c_1$ and
$a_1$ and then trace  on the resulting index in equation (N.12) then we
find that
$w_{d  a,}{}^{d}=\partial_\mu (det e_{a}{}^\mu )=0$. As has just  been
remarked   equation (N.13) only holds up to an inhomogeneous local Lorentz
transformations, but unfortunately $w_{d  a_1,}{}^{d}$ is  not of this
form in general. Thus although equation
(N.13) has many of the correct features it does not as it stands describe
the correct equation for gravity.   One might wonder if there is any 
term involving the dual gravity field  that one could add and that would
prevent
$w_{d  a_1,}{}^{d}$ being zero. However, such a term
would have to be constructed from the epsilon symbol, contain one
derivative acting on  the field
$h_{a_1\ldots a_8, b}$ and have, after the contraction,  only one index; 
one can easily see that this is not possible. 
\par
There are several ways out of this dilemma. The above computation could be
wrong in its details and in particular if the magnitude of the second term
in equation (N.13) was instead given by $-{7\over 15}$ then the terms
involving the usual gravity field would also be traceless.

Another possibility is that
there is some non-trivial dependence on the extra coordinates that leads
to a non-trivial trace.  However, the most likely possibility is that the
trace can be non-zero if one includes the contribution of the higher
level $E_{11}$ fields. The gravity fields  of the $E_{11}\otimes_s l_1$
non-linear realisation  occur at level
$0,3,6,9,\ldots $ and there  is in fact no reason to believe that the
full gravity equation arises only at the levels zero and three computed
above. In fact at level six we find the fields 
$$
h_{10,6,2},\  h_{9,8,1},\  h_{11,4,3},\ 
h_{11,5,1,1},\ h_{10,7,1},\ h_{11,6,1},\ h_{10,8},\ h_{11,7}
\eqno(N.14)$$
To contribute to the gravity equation we need an object with three
indices once we have differentiated with respect to the usual space-time. 
As this leads to an odd number of indices we cannot use a single epsilon.
It is very easy to find possible terms that have a trace, for example 
$$
\partial_{a}h^{b_1\ldots b_8 b}{}_{, b_1\ldots b_8}{}_{ , c}
\eqno(N.14)$$
The presence of this term would imply that the duality relation included
also higher level Cartan forms on its right-hand side, however, in view
of the non-linear nature of gravity this would not be unnatural.  
\par
This is the first time that the equation involving the gravity fields has
been systematically derived from the $E_{11}\otimes _s l_1$ non-linear
realisation. To derive this result we have only assumed that the 
 equation is linear in
derivatives and we have  set to zero  one constant; indeed the calculation
is  just a matter of $E_{11}$  group theory.  
 One cannot help be encouraged by
the  very intricate way in which the invariance is achieved  and it is
a convincing sign,  at least to
this  author, that this equation has almost all the correct features at
low levels.

\medskip
{\bf Additional  references given in this note added}
\medskip
\item {[N1]} T. Curtwright, {\it Generalised Gauge Fields}, Phys.Lett.
{\bf B165} (1985)  304. 
\item{[N2]} E. Bergshoeff, M de Roo, S. Kerstan, A.
Kleinschmidt, F. Riccioni, {\it  Dual Gravity and Matter}, 
   Gen.Rel.Grav.41 (2009) 39-48, arXiv:0803.1963.  

\item{[N3]} E. Bergshoeff, M de Roo, O, Hohm, {\it  Can dual gravity be
reconciled with E11? },  Phys.Lett.{\bf B675} (2009) 371-376,
arXiv:0903.4384.

\end